\documentclass[sigconf,natbib=true]{acmart}

\usepackage{multirow}
\usepackage{float}
\usepackage{url}
\usepackage{algorithm}
\usepackage{algpseudocode}
\usepackage{bm}
\usepackage{graphicx}
\usepackage{caption}
\usepackage{colortbl}
\usepackage[skip=0cm,list=true,labelfont=it]{subcaption}

\theoremstyle{definition}

\AtBeginDocument{%
  \providecommand\BibTeX{{%
    \normalfont B\kern-0.5em{\scshape i\kern-0.25em b}\kern-0.8em\TeX}}}

\copyrightyear{2025} 
\acmYear{2025} 
\setcopyright{cc}
\setcctype{by}
\acmConference[RecSys '25]{Proceedings of the Nineteenth ACM Conference on Recommender Systems}{September 22--26, 2025}{Prague, Czech Republic}
\acmBooktitle{Proceedings of the Nineteenth ACM Conference on Recommender Systems (RecSys '25), September 22--26, 2025, Prague, Czech Republic}\acmDOI{10.1145/3705328.3748082}
\acmISBN{979-8-4007-1364-4/2025/09}

\begin{document}

\title{Leave No One Behind: Fairness-Aware Cross-Domain Recommender Systems for Non-Overlapping Users}

\author{Weixin Chen}
\affiliation{
  \institution{Hong Kong Baptist University}
  \city{Hong Kong}
  \country{China}
}
\email{cswxchen@comp.hkbu.edu.hk}
\authornote{Equal contributions.}

\author{Yuhan Zhao}
\affiliation{
  \institution{Hong Kong Baptist University}
  \city{Hong Kong}
  \country{China}
}
\email{csyhzhao@comp.hkbu.edu.hk}
\authornotemark[1] 

\author{Li Chen}
\affiliation{
  \institution{Hong Kong Baptist University}
  \city{Hong Kong}
  \country{China}
}
\email{lichen@comp.hkbu.edu.hk}

\author{Weike Pan}
\affiliation{
  \institution{Shenzhen University}
  \city{Shenzhen}
  \country{China}
}
\email{panweike@szu.edu.cn}

\begin{abstract}
Cross-domain recommendation (CDR) methods predominantly leverage overlapping users to transfer knowledge from a source domain to a target domain. However, through empirical studies, we uncover a critical bias inherent in these approaches: while overlapping users experience significant enhancements in recommendation quality, non-overlapping users benefit minimally and even face performance degradation. This unfairness may erode user trust, and, consequently, negatively impact business engagement and revenue. To address this issue, we propose a novel solution that generates virtual source-domain users for non-overlapping target-domain users. Our method utilizes a dual attention mechanism to discern similarities between overlapping and non-overlapping users, thereby synthesizing realistic virtual user embeddings. We further introduce a limiter component that ensures the generated virtual users align with real-data distributions while preserving each user’s unique characteristics. Notably, our method is model-agnostic and can be seamlessly integrated into any CDR model. Comprehensive experiments conducted on three public datasets with five CDR baselines demonstrate that our method effectively mitigates the CDR non-overlapping user bias, without loss of overall accuracy. Our code is publicly available at \url{https://github.com/WeixinChen98/VUG}.

\end{abstract}

\ccsdesc[500]{Information systems}
\ccsdesc[500]{Information systems~Recommender systems}

\keywords{Cross-domain recommendation, fairness}
\maketitle

\section{INTRODUCTION}
Cross-domain recommendation (CDR) has emerged as a promising solution to provide better recommendations in the target domain with the help of the source domain~\cite{LCL24,XLW22,CSC22}. The central challenge of CDR lies in identifying and transferring suitable knowledge from the source domain to the target domain~\cite{ZWC21,ZZZ23}. 
Most of existing CDR methods utilize overlapping users to connect distinct domains, facilitating the mutual exchange of information~\cite{ZZH23,MSJ17}.
For example, BiTGCF~\cite{LLL20} incorporates cross-domain knowledge transfer into high-order connectivity in user-item graphs via the bridge of overlapping users.
\begin{figure}
    \centering
    \includegraphics[width=1\linewidth]{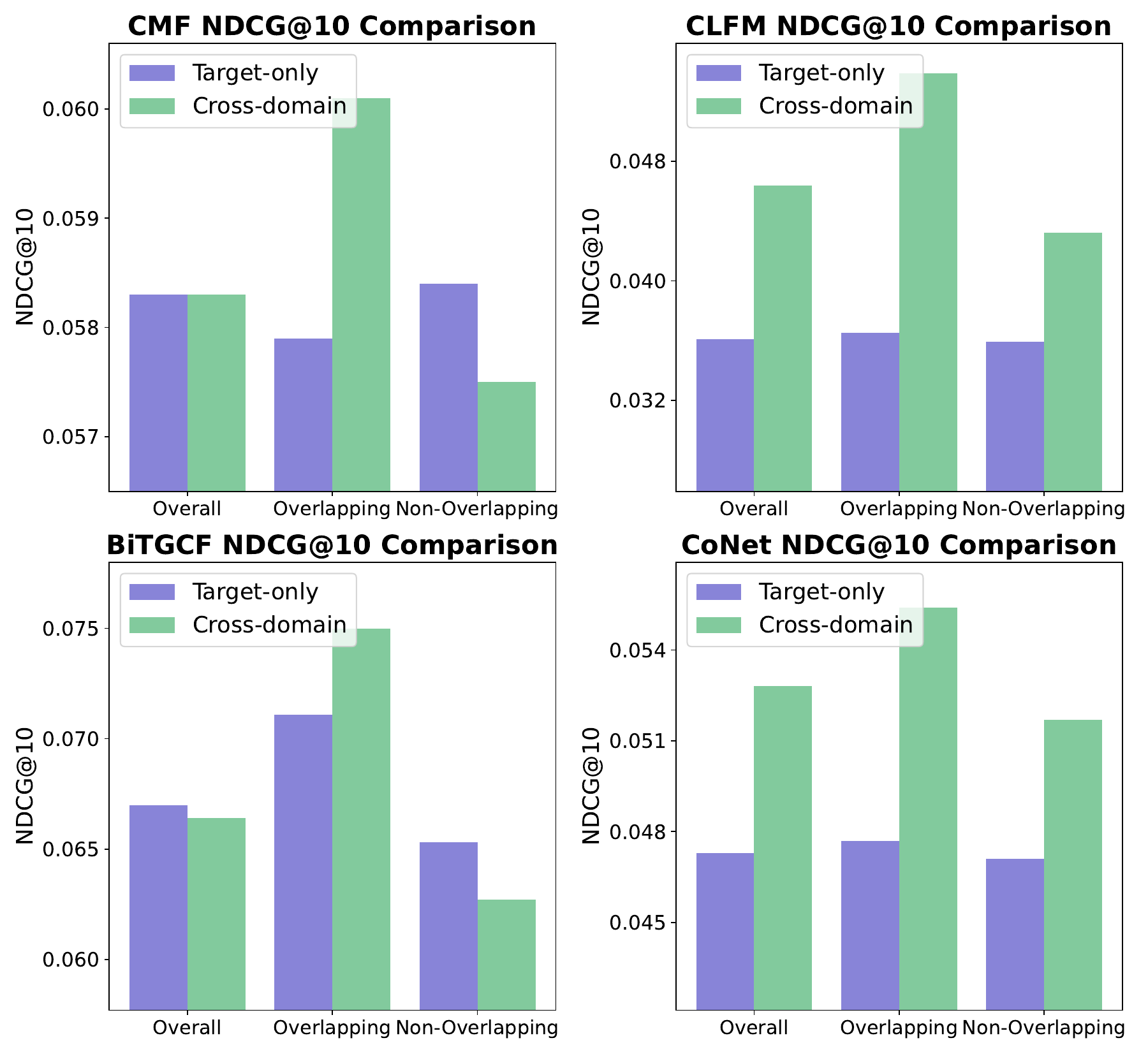}
    \caption{The performance comparison between overlapping users and non-overlapping users in the Epinions dataset. Target-only mode denotes the same method but disables the flow of cross-domain information.}
    \label{fig:intui}
\end{figure}

Despite the promising performance of those approaches, a critical fairness issue threatens their apparent success: \textit{they mainly focus on enhancing recommendations for overlapping users, while neglecting non-overlapping users.} As depicted in Figure~\ref{fig:intui}, while overlapping users obtain considerable improvement in recommendation quality, non-overlapping users experience negligible benefits or even suffer performance degradation.

This phenomenon is expected, as current methods prioritize overlapping users as bridges for information transfer, thereby likely marginalizing non-overlapping users. We term this issue the \textbf{CDR Overlapping Bias}: \textit{overlapping users receive the majority of benefits, whereas non-overlapping users receive minimal advantages or even experience degradation.}
In addition to these experimental findings, we also provide a theoretical analysis in Sec. ~\ref{sec:CDR_overlapping_bias_analysis} that further explains and validates this bias phenomenon from an information-theoretic perspective.
This disparity casts a shadow over the otherwise promising prospects of CDR. From a user perspective, perceived inequity can undermine trust, as users may feel discriminated. Furthermore, this imbalance poses challenges for companies seeking to boost engagement among non-overlapping users and ultimately impacts revenue. As a result, it may lead to a lose-lose situation for both non-overlapping users and business.

To address this problem, we propose a \textbf{\underline{v}}irtual \textbf{\underline{u}}ser \textbf{\underline{g}}eneration approach (referred to \textbf{VUG}) that creates synthetic (virtual) user profiles in the source domain for target-domain non-overlapping users. By doing so, we effectively transform non-overlapping users into (virtual) overlapping users, enabling them to harness the same cross-domain transfer benefits.
Moreover, once generated, virtual users can be seamlessly integrated into mainstream CDR pipelines, ensuring broad compatibility and a model-agnostic design. 

Implementing this idea faces two key challenges:

\begin{itemize} 
    \item How to generate virtual users for non-overlapping target users with no existing data in the source domain?
    \item How to ensure that the virtual users generated remain representative of the real source-domain data distribution?
\end{itemize}

To tackle the first challenge, we propose a generator based on the attention mechanism~\cite{ZBS18,LZW22}. Our approach identifies overlapping users in the target domain who are similar to non-overlapping users. Consequently, we generate virtual users in the source domain based on these similarities. Specifically, we use user-user relationships and item-item relationships in the source domain to determine attention weights, which are then used in the target domain to aggregate overlapping user data, resulting in the final virtual user.
For the second challenge, we introduce a limiter to ensure that the generated virtual user closely matches the real distribution of the source domain. During the training process, we treat overlapping users as a supervision signal, guiding the alignment of generated virtual embeddings toward the true source distribution. Furthermore, by leveraging a contrastive learning perspective~\cite{WYM22,WI20,ZCL}, we ensure that generated virtual users maintain distinct features. 

Our main contributions are summarized as follows:
\begin{itemize} 
    \item We empirically reveal the unfairness problem for non-overlapping users in cross-domain recommendation scenarios. 
    In addition, we theoretically analyze the causes of this unfairness from an information-theoretic perspective.
    \item We propose a cross-domain user generation framework that creates virtual users in the source domain for target-domain non-overlapping users. Our dual attention mechanism operates on levels of user similarity and interactive item similarity to derive attention weights, using overlapping users as a foundation to generate realistic virtual users. 
    \item We introduce a limiter to ensure that generated virtual user representations capture the characteristics of the source domain. By employing overlapping users as a supervision signal, we prevent the generated virtual embeddings from deviating from the source domain. Additionally, we attempt to preserve unique characteristics in the generated virtual users via contrastive learning. 
    \item We conduct extensive experiments on three public datasets, with five mainstream CDR baselines, demonstrating that our approach not only effectively mitigates CDR overlapping bias but also boosts overall recommendation accuracy. 
\end{itemize}

\section{CDR Overlapping Bias}
\subsection{Cross-Domain Recommendation}
In this work, we explore a general CDR scenario involving two domains: a source domain $\mathcal{D}^S$ and a target domain $\mathcal{D}^T$. The source domain is characterized by rich and informative interactions, whereas the target domain is relatively sparse. Notably, there exists a subset of overlapping users $\mathcal{U}^{o}$ who are present in both domains.

Each domain has its own set of users ($\mathcal{U}^{S}$ and $\mathcal{U}^{T}$), items ($\mathcal{I}^{S}$ and $\mathcal{I}^{T}$), and interaction records ($\mathcal{R}^{S}$ and $\mathcal{R}^{T}$). Given the observed data from both $\mathcal{D}^T$ and $\mathcal{D}^S$, CDR methods initially employ an embedding layer to derive embedding tables $\textbf{E}^{T}_u$ and $\textbf{E}^{S}_u$. Specifically, for a user $u \in \mathcal{U}^{o}$, there exist user embeddings $\textbf{e}^{S}$ in the source domain and $\textbf{e}^{T}$ in the target domain, while other users possess embeddings solely within their respective domains.

The parameter $\Theta$ training objective is to enhance recommendation performance in the target domain through techniques such as transfer learning~\cite{ZWC21, ZLZY23}, formulated as:
\begin{equation}
\label{eq:preliminary_cdr}
\begin{aligned}
    \underset{\Theta}{\text{maximize}} \quad P_{\mathcal{D}^T}\bigl(\Theta \mid \mathcal{D}^T,\mathcal{D}^S\bigr).
\end{aligned}
\end{equation}

\subsection{CDR Overlapping Bias Analysis}
\label{sec:CDR_overlapping_bias_analysis}
We adopt an information-theoretic perspective to explain why using overlapping users as a bridge between source and target domains may induce bias. Consider the source domain \(\mathcal{D}^S\) and target domain \(\mathcal{D}^T\) as source and destination in information theory. The CDR method we construct serves as the channel between them.

For overlapping users $\mathcal{U}^o$, we observe records $\mathcal{R}^S_u$ and $\mathcal{R}^T_u$ in the source and target domains, respectively.  While these records manifest differently due to domain-specific characteristics, they originate from the same underlying person.  We therefore posit the existence of a shared latent variable $\mathbf{z}_u$ that influences the record generation process in both domains:

\begin{equation}
\mathcal{R}^S_u \sim p_S(r \mid \mathbf{z}_u), \quad \mathcal{R}^T_u \sim p_T(r \mid \mathbf{z}_u).
\end{equation}

It is worth mentioning that other factors undoubtedly influence the record generation process, we assume their effects are consistent across both overlapping and non-overlapping users, and thus omit them for clarity in our analysis. And, for brevity, we will omit the subscript $u$. The joint probability of \(\mathcal{R}^S\) and \(\mathcal{R}^T\) is 
\begin{equation}
p(\mathcal{R}^S, \mathcal{R}^T)
= \int p(\mathbf{z})\,p_S(\mathcal{R}^S \mid \mathbf{z})\,p_T(\mathcal{R}^T \mid \mathbf{z})\, d \mathbf{z},
\end{equation}
where \(p(\mathcal{R}^S,\mathcal{R}^T)\) is not factorable as \(p(\mathcal{R}^S) \, p(\mathcal{R}^T)\). Therefore,
the mutual information satisfies:
\begin{equation}
I(\mathcal{R}^S; \mathcal{R}^T) 
= D_{\text{KL}}\bigl(p(\mathcal{R}^S,\mathcal{R}^T)\,\|\,p(\mathcal{R}^S)p(\mathcal{R}^T)\bigr) > 0.
\end{equation}
By contrast, for non-overlapping users, there is no shared $\mathbf{z}$  connecting \(\mathcal{R}^S\) and \(\mathcal{R}^T\), implying
\begin{equation}
I(\mathcal{R}^S; \mathcal{R}^T) = 0.
\end{equation}

The conditional entropy of \(\mathcal{R}^T\) given \(\mathcal{R}^S\) thus differs between overlapping and non-overlapping users:
\begin{equation}
H(\mathcal{R}^T \mid \mathcal{R}^S) 
= 
\begin{cases}
H(\mathcal{R}^T) - I(\mathcal{R}^S;\mathcal{R}^T) < H(\mathcal{R}^T), & \text{(overlapping)}\\
H(\mathcal{R}^T), & \text{(non-overlapping).}
\end{cases}
\end{equation}
By Fano's Inequality, the lower bound on the prediction error probability \(P_e\) for the target domain becomes:
\begin{equation}
P_e \geq \frac{H(\mathcal{R}^T \mid \mathcal{R}^S) - 1}{\log |\mathcal{V}^T|},
\end{equation}
where $|\mathcal{V}^T|$ denotes the cardinality of $\mathcal{R}^T$. Since overlapping users reduce \(H(\mathcal{R}^T\mid\mathcal{R}^S)\), they enjoy strictly lower prediction error bounds and typically benefit more from CDR. This advantage can, however, lead to a fairness concern, where non-overlapping users may experience less performance gain.

\subsection{Fairness Objective}
To quantify this disparity, we introduce a common fairness indicator, user-oriented group fairness (UGF)~\cite{UGF}. 
UGF in our study is formally defined as follows:
\begin{equation}
\label{eq:user_oriented_unfairness}
U G F =\left|\frac{1}{\left|\mathcal{U}^{o}\right|} \sum_{u \in \mathcal{U}^{o}} \mathcal{M}\left(L_{u}\right)-\frac{1}{\left|\mathcal{U}^{T} \setminus \mathcal{U}^{o}\right|} \sum_{u \in \mathcal{U}^{T} \setminus \mathcal{U}^{o}} \mathcal{M}\left(L_{u}\right)\right|,
\end{equation}
where $\mathcal{M}(L_u)$ is a metric that evaluates recommendation quality (e.g., NDCG and Hit Rate) for user $u$. Zero UGF signifies that the recommendations of equal quality are offered to different groups. 

In CDR, the aim could be to maximize the overall accuracy while limiting the UGF not larger than $\varepsilon$ as the strictness of fairness requirements between overlapping and non-overlapping users:
\begin{equation}
\label{eq:ugf}
\begin{aligned}
    &\underset{\Theta}{\text{maximize}} \quad P_{\mathcal{D}^T}\bigl(\Theta \mid \mathcal{D}^T,\mathcal{D}^S\bigr), \\
    &\text{subject to} \quad UGF \leq \varepsilon.
\end{aligned}
\end{equation}
In practice, however, most approaches focus on directly minimizing this metric, rather than explicitly specifying the constraint.
\section{Methodology}
\begin{figure*}
    \centering
    \includegraphics[width=0.95\linewidth]{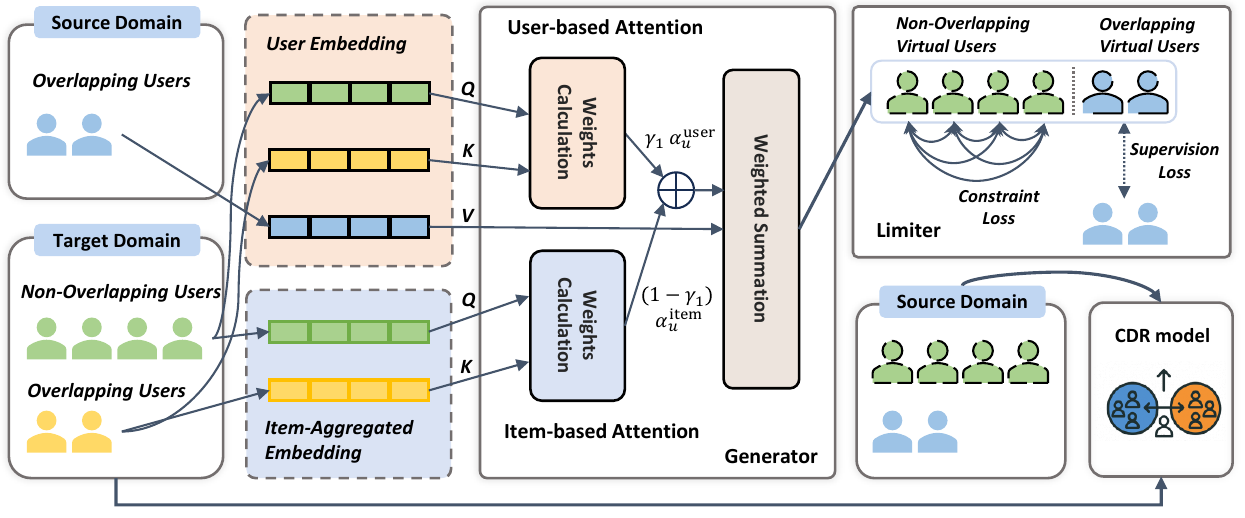}
    \caption{The illustration of our proposed VUG.}
    \label{fig:model_illustration}
\end{figure*}

To address the unfairness issue, we propose a novel \textit{virtual user generation} approach. Specifically, for every non-overlapping user in the target domain, we generate a corresponding virtual user in the source domain. Our approach is underpinned by two key components: (1) a \textit{generator} that creates virtual users, and (2) a \textit{limiter} that ensures the generated virtual users accurately reflect the characteristics of the source domain. The overall workflow is depicted in Figure~\ref{fig:model_illustration}. Importantly, our method generates virtual users without altering the original CDR framework, making it \textit{model-agnostic} and broadly applicable to various CDR algorithms to enhance fairness.


\subsection{Generator}

While the idea of generating virtual users in the source domain for non-overlapping users in the target domain is conceptually straightforward, its implementation poses significant challenges due to the absence of available data about non-overlapping users in the source domain.

While sophisticated generative models such as diffusion models~\cite{CHI23} and variational autoencoders (VAEs)~\cite{CLJH22} could be employed, these approaches introduce significant complexity, potentially impacting the efficiency of the original CDR method and posing challenges in training and design. Therefore, we propose a more straightforward and intuitive solution. For a given non-overlapping user $u_{\text{non}}$, we first identify the top-$N$ most behaviorally similar overlapping users within the target domain.
Recent research~\cite{ZJK16,HC10} reveals that these users are more likely to exhibit similar behavior to $u_{non}$ in the source domain. Based on this idea, we define:
\begin{align}
\{u_1, u_2, \dots, u_N\} &= \underset{u \in \mathcal{U}^{o}}{\arg \, {top}N} \; Sim(\mathbf{e}_{non}^{T}, \mathbf{e}^{T}_{u}), \\
\mathbf{e}_{non}^{S^\prime} &= Aggr(\mathbf{e}_{u}^{S}), \quad u \in \{u_1, u_2, \dots, u_N\},
\end{align}
where $\mathbf{e}_{non}^{T}$ denotes the embedding of the non-overlapping user $u_{non}$ in the target domain, $Sim(\cdot, \cdot)$ computes the behavior similarity between different users, and $Aggr(\cdot)$ represents an aggregation function (e.g., mean). After identifying the top-$N$ overlapping users, their embeddings in the source domain are aggregated to construct the virtual user embedding as the representation in the source domain for non-overlapping target-domain user $u_{non}$.

However, this approach has two critical limitations: (1) How to effectively identify behavior-similar users. (2) The top-$N$ filtering scheme is non-differentiable. 
To overcome these issues, we reformulate the process using an novel dual attention mechanism~\cite{ZCH24}. The final virtual user embedding is computed as:
\begin{equation}
\mathbf{e}_{non}^{S^\prime} \;=\;
\sum_{u \in \mathcal{U}^{o}}
\alpha_{u}\,\bigl(\mathbf{W}_{v}\,\mathbf{e}_{u}^{S} + \mathbf{b}_{v} \bigr),
\end{equation}
where $\mathbf{W}_{v} \in \mathbb{R}^{d \times d}$ is a trainable matrix, $\mathbf{b}_{v} \in \mathbb{R}^{d}$ is a bias term, and $\alpha_{u}$ represents the attention weight between the non-overlapping user and each overlapping user. 

This idea is that we consider that all overlapping users are likely to be behavior-similar users, and the likelihood probability is learned by attention weight. Then, we aggregate the embeddings $\mathbf{e}_{u}^{S}$ of overlapping users using learned attention weights to generate virtual users. In this way, users with a high probability of being behavior-similar users will contribute more to the virtual user, while users with low probability will do the opposite.

To calculate the attention weight, inspired by traditional collaborative filtering techniques~\cite{ZCH24,WDR06,chen2022global}, such as user-based kNN and item-based kNN, we compute $\alpha_{u}$ by considering both user-user and item-item relationships to capture the collaborative signals:
\begin{equation}
\alpha_{u} \;=\; \gamma_{1}\,\alpha_{u}^{\text{user}} \;+\; (1 - \gamma_{1})\,\alpha_{u}^{\text{item}},
\end{equation}
where $\alpha_{u}^{\text{user}}$ captures the user-user similarity, and $\alpha_{u}^{\text{item}}$ captures the item-item similarity. $\gamma_{1}$ is a hyperparameter to control the relative weights.

The user-based attention weight $\alpha_{u}^{\text{user}}$ is computed as:
\begin{align}
\alpha_{u}^{\text{user}} &= \frac{\exp\left(\beta_u^{\text{user}}\right)}{\sum_{{u^\prime} \in \mathcal{U}^o} \exp\left(\beta_{u^\prime}^{\text{user}}\right)}, \\
\beta_{u}^{\text{user}} &= 
\frac{ (\mathbf{W}_{q}^{\text{user}}\,\mathbf{e}_{non}^{T} + \mathbf{b}^{\text{user}}_{q}) (\mathbf{W}_{k}^{\text{user}}\,\mathbf{e}_{u}^{T} + \mathbf{b}^{\text{user}}_{k})^{\top}}{\sqrt{d}}
,
\end{align}
where $\mathbf{W}_{q}^{\text{user}}, \mathbf{W}_{k}^{\text{user}} \in \mathbb{R}^{d \times d}$ and $\mathbf{b}^{\text{user}}_{q}, \mathbf{b}^{\text{user}}_{k} \in \mathbb{R}^{d}$ are trainable parameters. Here, $\mathbf{e}_{non}^{T}$ serves as the query vector, and $\mathbf{e}_{u}^{T}$ ($\forall u \in \mathcal{U}^{o}$) acts as the key vector.

The item-based attention weight $\alpha_{u}^{\text{item}}$ measures the similarity between overlapping and non-overlapping users based on their interacted items. First, for each user $u$ in target domain, we compute the aggregated item embedding $\mathbf{g}_{u}^{T}$:
\begin{align}
\mathbf{g}_{u}^{T} = \frac{1}{|\mathcal{I}^T_u|} \sum_{i \in \mathcal{I}^T_u} \mathbf{e}_{i}.
\end{align}
This step considers the interactions between users and items holistically, rather than focusing on individual items. The rationale is that we are not concerned with the impact of any specific interaction; instead, our focus lies on capturing the overall similarity. The item-based attention weight is calculated similarly to user-based attention, but using item aggregated embeddings as queries and keys:
$\alpha_{u}^{\text{item}}$ is computed as:
\begin{align}
\alpha_{u}^{\text{item}} &= \frac{\exp\left(\beta_u^{\text{item}}\right)}{\sum_{u^\prime \in \mathcal{U}^o} \exp\left(\beta_{u^\prime}^{\text{item}}\right)}, \\
\beta_{u}^{\text{item}} &= 
\frac{ (\mathbf{W}_{q}^{\text{item}}\,\mathbf{g}_{non}^{T} + \mathbf{b}^{\text{item}}_{q}) (\mathbf{W}_{k}^{\text{item}}\,\mathbf{g}_{u}^{T} + \mathbf{b}^{\text{item}}_{k})^{\top}}{\sqrt{d}}.
\end{align}
In summary, for $\alpha_{u}^{\text{user}}$, we use the non-overlapping user's embedding in the target domain as the query and the overlapping user's embedding as the key, with the overlapping user's source-domain embedding as the value. For $\alpha_{u}^{\text{item}}$, we use the aggregated item embeddings of the non-overlapping and overlapping users as queries and keys, respectively, with the overlapping user's source-domain embedding as the value.

This attention-based approach effectively aggregates overlapping user information from the source domain to construct virtual users. The entire process is fully differentiable and addresses the aforementioned challenges.

\subsection{Limiter}

Although the aforementioned solution generates a virtual user, we recognize two critical challenges that need to be addressed:

\begin{itemize}
    \item The source domain inherently possesses its own data distribution. How can we ensure that the generated users align with the characteristics of the source domain?
    \item 
    Even if two users are friends in the source domain, they may display their own
    distinct interest in the target domain. How can we preserve the identity of the user as much as possible while accounting for these differences?
\end{itemize}

To address the first challenge, we propose introducing a supervision signal that guides the generator to produce user embeddings that closely match the characteristics of the source domain. A natural question arises: Where does this supervisory signal come from? To solve this, we shift our eyes to overlapping users—those who exist in both the source and target domains. If the generator takes the embedding of an overlapping user from the target domain as input, the output should ideally approximate the embedding of the same user in the source domain. We can leverage them as a supervisory signal by encouraging the generator to minimize the discrepancy between the generated and true embeddings.

To operationalize this idea, we first pass the embeddings of overlapping users through the generator to obtain their corresponding representations in the source domain:
\begin{align}
    \mathbf{e}_{o}^{S^\prime} = \text{generator}(\mathbf{e}_{o}^{T}),
\end{align}
where $\mathbf{e}_{o}^{T}$ denotes the embedding of an overlapping user in the target domain, and $\mathbf{e}_{o}^{S^\prime}$ represents the generated embedding in the source domain.

Next, we enforce the generated embedding to closely approximate the true embedding by minimizing the following MSE loss:
\begin{align}
\mathcal{L}_{\text{super}} = \frac{1}{|\mathcal{U}^o|}\sum_{o \in \mathcal{U}^o} \| \mathbf{e}_{o}^{S^\prime} - \mathbf{e}_{o}^{S} \|^{2},
\end{align}
where $\mathbf{e}_{o}^{S}$ is the true embedding of the overlapping user in the source domain, and $\mathcal{U}^{o}$ denotes the set of overlapping users. Here, we employ the Euclidean distance to measure the similarity between embeddings, a choice motivated by its simplicity, interpretability, and computational efficiency~\cite{ZCL23}. This distance metric is widely used in collaborative filtering for embedding approximation tasks.

As previously discussed, this supervision loss is specifically designed to guide the generator. Therefore, during the optimization process, we freeze the parameters of other components to ensure that the generator is the sole component being updated~\cite{ZCC24,ZS23}.
\begin{align}
    \min_{\Theta_{\text{gen}}} \mathcal{L}_{\text{super}},
\end{align}
where $\Theta_{\text{gen}}$ represents the parameters of the generator.

As for the second challenge, we mentioned earlier that users who are similar in the target domain are more likely to be similar in the source domain than other users. However, these users are not all the same, and they may have their unique interests and behaviors. Therefore, we want to design a loss function to constrain virtual users not to completely copy the existing information of overlapping users, but to retain their unique characteristics. 

Our approach draws inspiration from contrastive learning (CL), a powerful technique for extracting latent representations from unlabeled data that has demonstrated considerable success in recommendation systems~\cite{ZPL23,YZL23}.  CL typically involves two key processes: alignment and uniformity~\cite{WYM22,WI20}. Alignment ensures that similar samples are mapped to nearby embeddings. Since our generator produces virtual user embeddings that already give greater weight to similar users more closely, we can remove the explicit constraint. Another process, uniformity, encourages embeddings to be evenly distributed on the unit hypersphere, thereby preserving as much intrinsic information about each data point as possible.

The principle of uniformity aligns perfectly with our goal of enabling generated virtual users to retain their individual characteristics. To achieve this, we introduce a constraint loss function defined as:
\begin{equation}
\mathcal{L}_{\text{constrain}} = \log \mathbb{E} \left[ e^{-2\|\mathbf{e}_{u}^{S^\prime}-\mathbf{e}_{u^\prime}^{S^\prime}\|^2} \right], \quad u, u^\prime \in \mathcal{U}^T \setminus \mathcal{U}^o
\end{equation}
where $\mathbf{e}_{u}^{S^\prime}$ and $\mathbf{e}_{u^\prime}^{S^\prime}$ represent the embeddings of two distinct virtual users in the source domain.
This loss encourages all generated virtual users to remain distant from one another, thereby preserving their unique identities. This loss and our original generator design can be seen as finding a balance, where we hope that the generated virtual users can learn knowledge from other overlapping users while maintaining their own unique interest.

Finally, the generated virtual users can be seamlessly integrated into the subsequent CDR recommendation pipeline. The overall training objective is split into the original CDR loss for the main model and the combined supervision and constraint losses for the generator, ensuring that each part is appropriately optimized. The complete optimization process is formulated as:
\begin{align}
    \min&_{\Theta \setminus \Theta_{\text{gen}}}  \ \mathcal{L}_{\text{CDR}}, \\
    \min&_{\Theta_{\text{gen}}}  \, \gamma_{2} \, \mathcal{L}_{\text{super}}  + (1-\gamma_{2}) \, \mathcal{L}_{\text{constrain}},
\end{align}
where $\mathcal{L}_{\text{CDR}}$ denotes the original CDR loss function, $\gamma_{2}$ is a hyperparameter controlling the trade-off between two losses, and $\Theta$ represents the set of all model parameters.

\section{EXPERIMENTS}



{We conduct comprehensive experiments to evaluate the performance of VUG. Specifically, we evaluate the effectiveness of integrating VUG in mitigating CDR overlapping bias across various CDR models (Sec.~\ref{sec:bias_mitigation}). We analyze the contribution of individual VUG components to model performance (Sec.~\ref{sec:component_analysis}) and investigate the impact of different parameter settings on VUG's efficacy (Sec.~\ref{sec:parameter_sensitivity}). Additionally, we examine how VUG performs under varying overlapping ratios (Sec.~\ref{sec:overlap_analysis}) and assess its efficiency for practical applications (Sec.~\ref{sec:efficiency_analysis}).}

\subsection{Experimental Setup}
\subsubsection{Datasets}
We conduct experiments on four datasets widely used in the literature to evaluate the performance of VUG: 
\begin{itemize} 
    \item \textbf{Amazon:} A comprehensive e-commerce dataset featuring user reviews across 24 domains, including Book and Movie from the e-commerce platform Amazon. The platform's extensive user community makes it a rich source for analytical research.
    \item \textbf{Douban:} Collected from the Douban platform, this dataset contains user reviews and interactions across three primary categories: Book, Movie, and Music.
    \item \textbf{Epinions:} Sourced from Epinions.com, this dataset encompasses consumer reviews spanning 587 domains and sub-domains, covering diverse product categories such as Book, Electronics, and Game.
\end{itemize}

\begin{table}
\centering
\caption{Statistics of the preprocessed cross-domain recommendation datasets used in our experiments. For each dataset, the domain in the first row is the source domain, and the other is the target domain. Overlap denotes the ratio of overlapped users over all users in the corresponding domain.}
\label{tab:statistics}
\begin{tabular}{llrrrr}
\toprule
\textbf{Dataset} & \textbf{Domain} & \textbf{\#Users} & \textbf{\#Items} & \textbf{\#Inter.} & \textbf{Overlap} \\
\midrule
\multirow{2}{*}{Amazon} & Book & 65,725 & 70,071 & 2,021,443 & 2.55\% \\
 & Movie & 8,663 & 7,790 & 229,257 & 19.32\% \\
\midrule
\multirow{2}{*}{Douban} & Book & 18,086 & 33,067 & 809,248 & 5.94\% \\
 & Movie & 3,372 & 9,342 & 311,797 & 31.88\% \\
\midrule
\multirow{2}{*}{Epinions} & Elec & 10,124 & 13,018 & 34,859 & 12.51\% \\
 & Game & 4,247 & 4,094 & 16,471 & 29.83\% \\
\bottomrule
\end{tabular}
\end{table}

\begin{table*}[t]
\caption{Experimental results of different CDR models with (w) or without (w/o) our VUG approach.
}
\centering
\begin{tabular}{@{}clllllllllllll@{}}
\toprule
\multirow{2}{*}{\textbf{Datasets}} & \multirow{2}{*}{\textbf{Metric}} & \multicolumn{2}{c}{\textbf{CMF}} & \multicolumn{2}{c}{\textbf{CLFM}} & \multicolumn{2}{c}{\textbf{BiTGCF}} & \multicolumn{2}{c}{\textbf{CoNet}} & \multicolumn{2}{c}{\textbf{MFGSLAE}} \\ \cmidrule(l){3-12} 
 &  & w/o & w & w/o & w & w/o & w & w/o & w & w/o & w \\ \midrule
\multirow{10}{*}{Amazon} 
 & HR@10 & 0.1063 & 0.1031 & 0.0582 & \textbf{0.0627} & 0.1041 & \textbf{0.1053} & 0.0500 & \textbf{0.0501} & 0.1048 & \textbf{0.1094} \\
 & HR@20 & 0.1578 & \textbf{0.1609} & 0.0972 & \textbf{0.1026} & 0.1593 & \textbf{0.1624} & 0.0820 & \textbf{0.0827} & 0.1583 & \textbf{0.1601} \\
 & NDCG@10 & 0.0356 & \textbf{0.0359} & 0.0177 & \textbf{0.0191} & 0.0354 & \textbf{0.0360} & 0.0150 & \textbf{0.0154} & 0.0373 & \textbf{0.0393} \\
 & NDCG@20 & 0.0455 & \textbf{0.0465} & 0.0241 & \textbf{0.0258} & 0.0451 & \textbf{0.0466} & 0.0199 & \textbf{0.0208} & 0.0469 & \textbf{0.0485} \\
\cmidrule(l){2-12} 
 & Accuracy Improvement &  & \textbf{+0.50\%} &  & \textbf{+7.06\%} &  & \textbf{+2.03\%} &  & \textbf{+2.06\%} &  & \textbf{+3.57\%}\\ \cmidrule(l){2-12} 
 & UGF (HR@10) & 0.0607 & \textbf{0.0573} & 0.0501 & \textbf{0.0363} & 0.0620 & \textbf{0.0538} & 0.0299 & \textbf{0.0208} & 0.0537 & \textbf{0.0347} \\
 & UGF (HR@20) & 0.0761 & \textbf{0.0672} & 0.0728 & \textbf{0.0491} & 0.0772 & \textbf{0.0667} & 0.0436 & \textbf{0.0397} & 0.0726 & \textbf{0.0555} \\
 & UGF (NDCG@10) & 0.0065 & \textbf{0.0058} & 0.0078 & \textbf{0.0018} & 0.0091 & \textbf{0.0040} & 0.0042 & \textbf{0.0003} & 0.0043 & \textbf{0.0018} \\
 & UGF (NDCG@20) & 0.0052 & \textbf{0.0033} & 0.0084 & \textbf{0.0007} & 0.0080 & \textbf{0.0033} & 0.0035 & \textbf{0.0001} & 0.0036 & \textbf{0.0024} \\
\cmidrule(l){2-12} 
 & Fairness Improvement &  & \textbf{+16.15\%} &  & \textbf{+57.17\%} &  & \textbf{+35.41\%} &  & \textbf{+57.34\%} &  & \textbf{+37.60\%} \\ \midrule
\multirow{10}{*}{Douban} 
 & HR@10 & 0.2696 & \textbf{0.2927} & 0.2284 & \textbf{0.2536} & 0.2281 & 0.2263 & 0.2553 & \textbf{0.2598} & 0.3256 & \textbf{0.3283} \\
 & HR@20 & 0.3493 & \textbf{0.3870} & 0.3004 & \textbf{0.3434} & 0.3028 & \textbf{0.3052} & 0.3375 & 0.3360 & 0.4039 & \textbf{0.4104} \\
 & NDCG@10 & 0.0822 & \textbf{0.0905} & 0.0633 & \textbf{0.0765} & 0.0656 & \textbf{0.0674} & 0.0767 & \textbf{0.0804} & 0.1061 & 0.1059 \\
 & NDCG@20 & 0.0901 & \textbf{0.1018} & 0.0694 & \textbf{0.0868} & 0.0729 & \textbf{0.0749} & 0.0843 & \textbf{0.0874} & 0.1155 & \textbf{0.1162} \\
\cmidrule(l){2-12} 
 & Accuracy Improvement &  & \textbf{+10.61\%} &  & \textbf{+17.82\%} &  & \textbf{+1.37\%} &  & \textbf{+2.45\%} &  & \textbf{+0.71\%}\\ \cmidrule(l){2-12} 
 & UGF (HR@10) & 0.2734 & \textbf{0.2517} & 0.2301 & \textbf{0.2095} & 0.2442 & \textbf{0.2386} & 0.2383 & \textbf{0.2331} & 0.2758 & 0.2883 \\
 & UGF (HR@20) & 0.3106 & \textbf{0.2949} & 0.2759 & \textbf{0.2742} & 0.2684 & 0.2717 & 0.2844 & \textbf{0.2742} & 0.2933 & 0.2947 \\
 & UGF (NDCG@10) & 0.0480 & \textbf{0.0403} & 0.0371 & \textbf{0.0305} & 0.0395 & \textbf{0.0368} & 0.0381 & \textbf{0.0372} & 0.0473 & \textbf{0.0435} \\
 & UGF (NDCG@20) & 0.0399 & \textbf{0.0299} & 0.0280 & \textbf{0.0245} & 0.0280 & \textbf{0.0273} & 0.0288 & \textbf{0.0273} & 0.0326 & \textbf{0.0324} \\
\cmidrule(l){2-12} 
 & Fairness Improvement &  & \textbf{+13.52\%} &  & \textbf{+9.96\%} &  & \textbf{+2.60\%} &  & \textbf{+3.33\%} &  & \textbf{+0.91\%} \\ \midrule
\multirow{10}{*}{Epinions} 
 & HR@10 & 0.1069 & \textbf{0.1077} & 0.0962 & \textbf{0.1126} & 0.1199 & \textbf{0.1348} & 0.1115 & 0.1100 & 0.1161 & \textbf{0.1363} \\
 & HR@20 & 0.1535 & \textbf{0.1615} & 0.1558 & \textbf{0.1665} & 0.1863 & \textbf{0.2008} & 0.1611 & \textbf{0.1737} & 0.1730 & \textbf{0.2001} \\
 & NDCG@10 & 0.0583 & \textbf{0.0589} & 0.0464 & \textbf{0.0546} & 0.0664 & \textbf{0.0689} & 0.0528 & 0.0514 & 0.0575 & \textbf{0.0710} \\
 & NDCG@20 & 0.0700 & \textbf{0.0721} & 0.0612 & \textbf{0.0681} & 0.0826 & \textbf{0.0851} & 0.0650 & \textbf{0.0670} & 0.0713 & \textbf{0.0865} \\
\cmidrule(l){2-12} 
 & Accuracy Improvement &  & \textbf{+2.50\%} &  & \textbf{+13.22\%} &  & \textbf{+6.75\%} &  & \textbf{+1.73\%} &  & \textbf{+19.47\%}\\ \cmidrule(l){2-12} 
 & UGF (HR@10) & 0.0122 & \textbf{0.0020} & 0.0384 & \textbf{0.0115} & 0.0266 & \textbf{0.0128} & 0.0259 & \textbf{0.0244} & 0.0230 & \textbf{0.0003} \\
 & UGF (HR@20) & 0.0135 & \textbf{0.0112} & 0.0377 & \textbf{0.0060} & 0.0106 & \textbf{0.0046} & 0.0374 & \textbf{0.0304} & 0.0297 & \textbf{0.0015} \\
 & UGF (NDCG@10) & 0.0026 & \textbf{0.0001} & 0.0107 & \textbf{0.0008} & 0.0123 & \textbf{0.0000} & 0.0037 & \textbf{0.0036} & 0.0099 & \textbf{0.0005} \\
 & UGF (NDCG@20) & 0.0029 & \textbf{0.0018} & 0.0103 & \textbf{0.0002} & 0.0079 & \textbf{0.0022} & 0.0058 & \textbf{0.0044} & 0.0116 & \textbf{0.0014} \\
\cmidrule(l){2-12} 
 & Fairness Improvement &  & \textbf{+58.68\%} &  & \textbf{+86.18\%} &  & \textbf{+70.16\%} &  & \textbf{+12.84\%} &  & \textbf{+94.13\%} \\ 
\bottomrule
\end{tabular}
\label{tab:main_results}
\end{table*}

Following~\cite{WYZ25}, we perform 5-core filtering to remove users and items with fewer than five interactions for the expansive datasets Douban
and Amazon. 
Table~\ref{tab:statistics} summarizes the used domains and the statistics of the datasets. 

\subsubsection{Implementation Details}
The experiments are conducted on single NVIDIA Tesla V100-32GB GPU using PyTorch. 
To ensure reproducibility, we implement all methods and apply pre-processing, dataset split, and evaluation using the RecBole CDR framework standard settings~\cite{ZMH21, ZHP22}.
{Specifically, an 8:2 split was used for training and validation in the source domain, and an 8:1:1 split for training, validation, and test in the target domain. Ratings of 3 and above were considered positive, with evaluations conducted across all items. The training, validation, and test splits were maintained separately by each user.}
We maintain a fixed size of 64 for the embeddings. 
The optimization of parameters is carried out using Adam~\cite{KB14} with a default learning rate of 0.001 and a default mini-batch size of 2048.
The $L_2$ regularization coefficient is set to $10^{-4}$ by default.
We evaluate the performance of VUG by grid search both $\gamma_{1}$ and $\gamma_{2}$ from 0 to 1 with a step size of 0.1. 
We meticulously tune all hyper-parameters on the validation sets and report the best performance for all baseline models.

\subsubsection{Evaluation Protocols}
To evaluate accuracy employing standard metrics for top-K recommendations, we encompass hit rate (HR@K) and normalized discounted cumulative gain (NDCG@K). For fairness measurement, we use UGF in Equation~\eqref{eq:ugf} equipped with above metrics to evaluate the inequality in recommendation quality between overlapping users and non-overlapping users. 



\subsubsection{Baseline}
To validate the effectiveness of our solution, we integrate VUG with a wide range of representative and state-of-the-art methods.
\begin{itemize}
    \item \textbf{CMF}~\cite{SG08} jointly factorizes multiple rating matrices by sharing latent parameters across different domains.
    \item \textbf{CLFM}~\cite{GLC13} utilizes a cluster-level latent factor model to learn both shared cross-domain and domain-specific knowledge.
    \item \textbf{BiTGCF}~\cite{LLL20}  exploits high-order connectivity within domains while using overlapping users as bridges for cross-domain knowledge transfer.
    \item  \textbf{CoNet}~\cite{HZY18} leverages cross-connection units to facilitate effective knowledge transfer between domains.
    \item \textbf{MFGSLAE}~\cite{WYZ25} employs a factor selection module with bootstrapping to distinguish between domain-specific and shared information.
\end{itemize}

\subsection{Overall Performance Comparison}
\label{sec:bias_mitigation}
We report the overall performance results in Table~\ref{tab:main_results}. Reported improvements represent average gains across all evaluation metrics and are statistically significant ($p < 0.05$) over the best-performing baseline(s) using a two-sided t-test.  Key observations include:
\begin{itemize}
\item With the help of VUG, all base models show significant performance improvements in UGF-related metrics. The performance improvement of MFGSLAE on Epinions is as high as 94.13\%. This demonstrates VUG's effectiveness in mitigating CDR overlapping bias by enabling non-overlapping users to benefit from the same training strategies as overlapping users. VUG promotes fairer recommendations and potentially enhances user satisfaction.

\item Beyond fairness, VUG also yields significant performance gains for all base models across most datasets and metrics.  CLFM, for instance, exhibits a 17.82\% improvement on Douban. This indicates that addressing CDR overlapping bias not only improves fairness but also enhances overall recommendation quality, achieving a win-win scenario for both users and the platform.

\item Traditional methods show poor performance of UGF-related metrics. This and our previous Figure~\ref{fig:intui} can corroborate each other, indicating that the traditional method with overlapping users as the medium generally has serious CDR overlapping bias, which brings unfair experience to the non-overlapping users.

\item It is interesting to observe that even the most SOTA methods (such as MFGSLAE), brought a large improvement in overall performance; the bias situation on some datasets was even worse than that of classical methods such as CMF, causing great discrimination against non-overlapping users. This underscores that simply optimizing for performance does not necessarily address fairness concerns, and dedicated strategies are crucial for mitigating bias.
\end{itemize}

\begin{figure*}[htbp]
    \centering
    \includegraphics[width=1\linewidth]{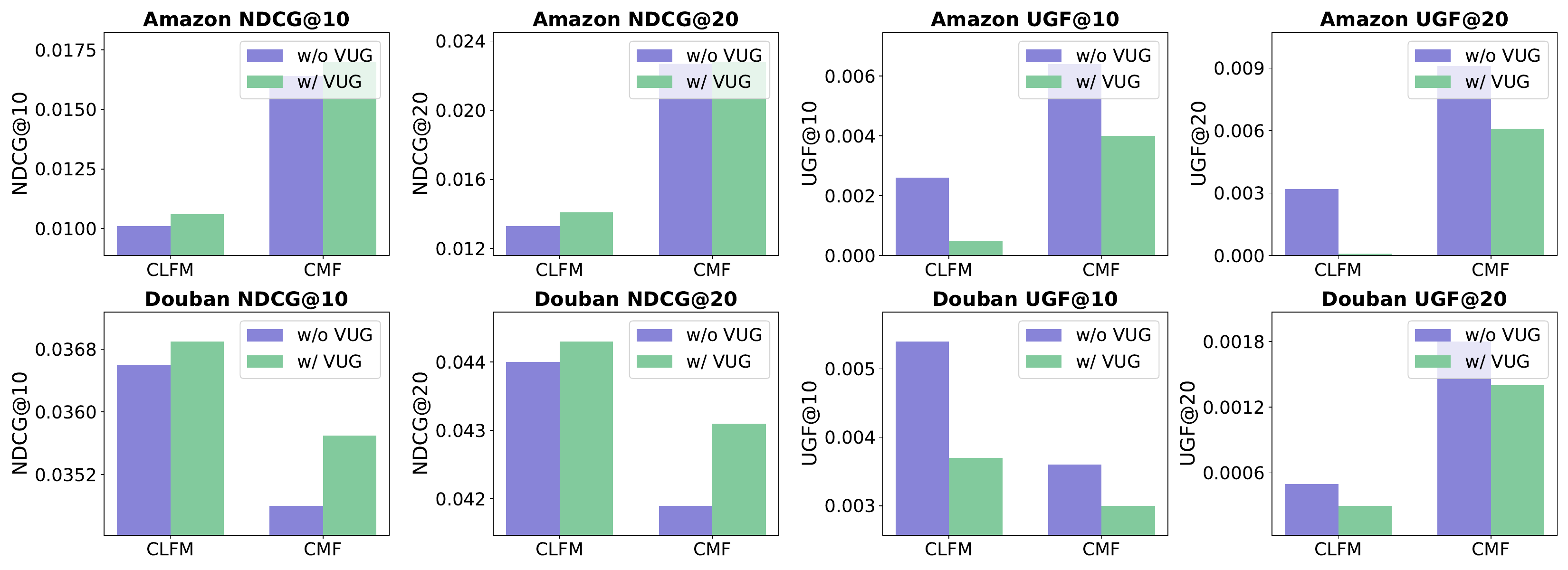}
    \caption{{Performance under temporal splitting on the Amazon and Douban datasets, for comparison with and without VUG.}}
    \label{fig:temporal_split}
\end{figure*}

{\textbf{Temporal split.} Figure~\ref{fig:temporal_split} compares CLFM and CMF with and without VUG under temporal splitting. Overall, VUG consistently improves both accuracy (NDCG) and fairness (UGF). These findings validate that our approach remains effective in more realistic, time-aware training scenarios. By synthesizing virtual cross-domain signals for non-overlapping users, VUG narrows the performance gap between overlapping and non-overlapping user groups, demonstrating its robustness across different data-splitting strategies.}



\subsection{Ablation Study}
\label{sec:component_analysis}
We analyze the effectiveness of different components via the following variants: (1) VUG without $\mathcal{L}_{\text{constrain}}$ (w/o $\mathcal{L}_\mathrm{constrain}$), (2) VUG without $\mathcal{L}_{\text{super}}$ (w/o $\mathcal{L}_{\text{super}}$), (3) VUG without user-user attention (w/o $\alpha_{u}^{\text{user}}$), and (4) VUG without item-item attention (w/o $\alpha_{u}^{\text{item}}$). 
The results are presented in Table~\ref{tab:ablation}. They suggest that all components positively contribute to model performance and fairness. 

It is worth noting that VUG without $\mathcal{L}_{\text{super}}$ has a large decline in the fairness metric. This is because this loss function constrains the generated virtual users to conform to the source domain distribution. Without this constraint, the virtual users could degenerate into noise, potentially hindering rather than improving the performance for non‐overlapping users.

\begin{table}[t]
\centering
\caption{Ablation study of our proposed method VUG versus its variants without specific components, integrated with MFGSLAE on the Epinions dataset.
}
\label{tab:ablation}
\begin{tabular}{lcccc}
\toprule
\multirow{2}{*}{\textbf{Method}} & \multicolumn{4}{c}{\textbf{Accuracy (\textit{larger} is better)}}\\
\cmidrule(lr){2-5}
& HR@10 & HR@20 & NDCG@10 & NDCG@20 \\ 
\midrule
\emph{w/o $\mathcal{L}_{\text{constrain}}$} & 0.1302 & 0.1974 & 0.0655 & 0.0821 \\ 
\emph{w/o $\mathcal{L}_{\text{super}}$} & 0.1355 & 0.1985 & 0.0707 & 0.0864 \\ 
\emph{w/o $\alpha_{u}^{\text{user}}$} & 0.1298 & 0.1974 & 0.0662 & 0.0828 \\ 
\emph{w/o $\alpha_{u}^{\text{item}}$} & 0.1287 & 0.1963 & 0.0658 & 0.0824 \\ 
\textbf{VUG} & \textbf{0.1363} & \textbf{0.2001} & \textbf{0.0710} & \textbf{0.0865} \\ 
\midrule
\multirow{2}{*}{\textbf{Method}} & \multicolumn{4}{c}{\textbf{UGF (\textit{smaller} is better)}}\\
\cmidrule(lr){2-5}
& HR@10 & HR@20 & NDCG@10 & NDCG@20 \\
\midrule
\emph{w/o $\mathcal{L}_{\text{constrain}}$} & 0.0010 & 0.0168 & 0.0005 & 0.0033 \\
\emph{w/o $\mathcal{L}_{\text{super}}$} & 0.0281 & 0.0152 & 0.0129 & 0.0096 \\
\emph{w/o $\alpha_{u}^{\text{user}}$} & 0.0020 & 0.0095 & 0.0015 & 0.0034 \\
\emph{w/o $\alpha_{u}^{\text{item}}$} & 0.0041 & 0.0112 & \textbf{0.0005} & 0.0035 \\
\textbf{VUG} & \textbf{0.0003} & \textbf{0.0015} & \textbf{0.0005} & \textbf{0.0014} \\
\bottomrule
\end{tabular}
\end{table}

\subsection{Hyperparameter Study}
\label{sec:parameter_sensitivity}
\begin{figure}[htbp]
    \centering
    \includegraphics[width=1\linewidth]{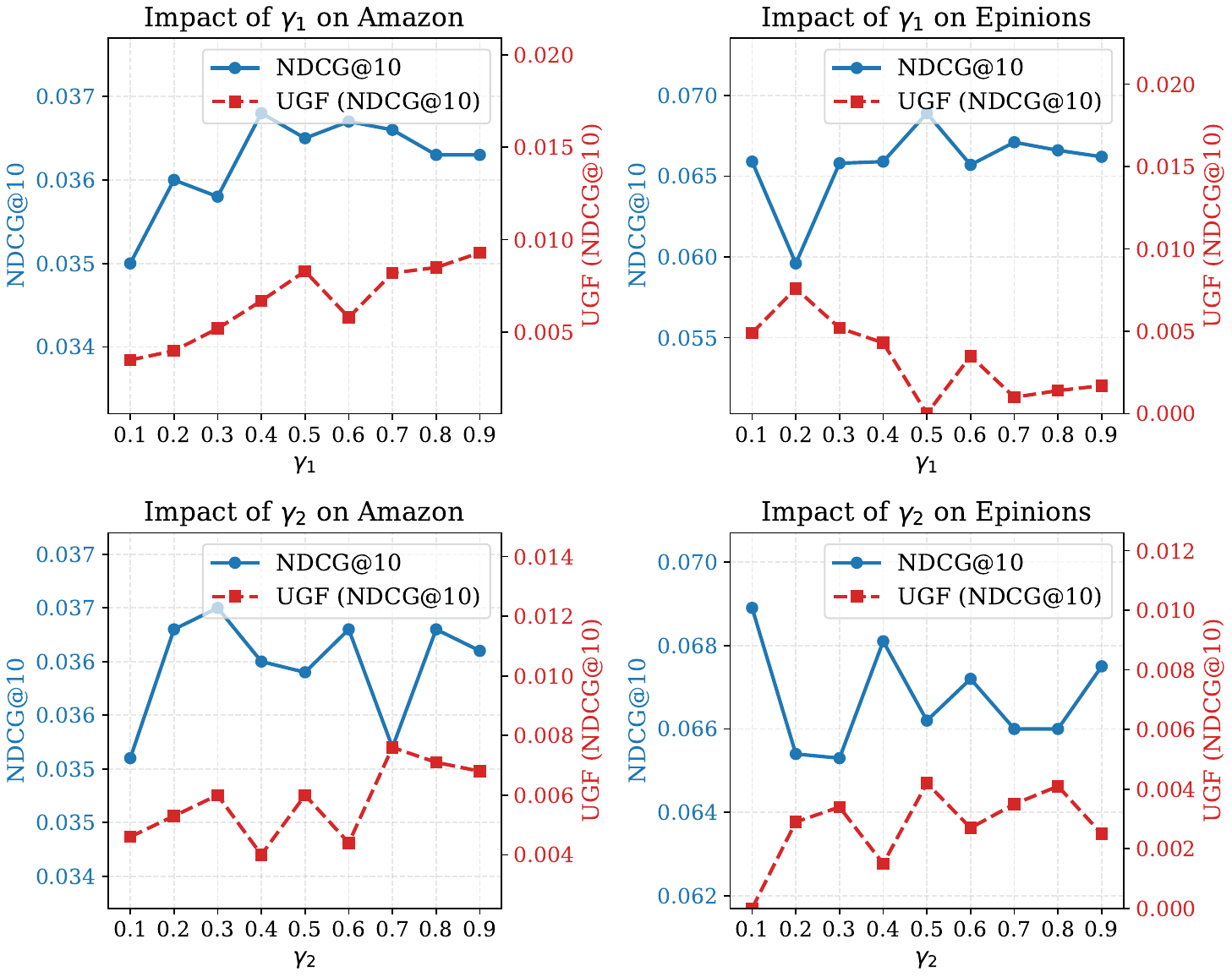}
    \caption{Impact of hyperparameters on VUG.}
    \label{fig:param_sensitivity}
\end{figure}

We conduct experiments on Amazon and Epinions, integrated with BiTGCF, to study the impact of different values of $\gamma_{1}$ and $\gamma_{2}$ and present the results in Figure~\ref{fig:param_sensitivity}. The parameter $\gamma_1$ balances the contributions of user-user and item-item attention weights.  As Figure~\ref{fig:param_sensitivity} illustrates, the model's performance exhibits a trend of initial improvement followed by a decline as $\gamma_1$ increases.  The optimal value of $\gamma_1$ varies across datasets, likely reflecting differences in the relative importance of user-user relationships within each dataset.  In some datasets, user-user relationships are more informative, while in others, item-item relationships may dominate.  Encouragingly, VUG demonstrates robust performance across a wide range of $\gamma_1$ values. The parameter $\gamma_2$ balances the influence of the supervised loss $\mathcal{L}_{\text{super}}$ and the constraint loss $\mathcal{L}_{\text{constrain}}$.  A small $\gamma_2$ may provide insufficient supervision for the generator, resulting in virtual users that do not conform to the source domain distribution and consequently reduced performance.  Conversely, a large $\gamma_2$ may over-emphasize the constraint, causing the generated virtual users to lose their distinct identities, potentially also hindering performance. However, VUG maintains strong performance across a wide range of $\gamma_2$ values.

\subsection{Overlapping Ratio Analysis}
\label{sec:overlap_analysis}

{In this section, we investigate the influence of varying overlap ratios on the performance of our proposed method VUG. }
{From Table~\ref{tab:overlap_ratio}, we observe that VUG consistently improves recommendation accuracy while substantially reducing the unfairness gap. Notably, the gains are more pronounced at lower overlap ratios, precisely the scenario in which non-overlapping users are at a greater disadvantage. 
The reduced UGF values demonstrate that VUG effectively narrows the disparity between overlapping and non-overlapping user groups, ensuring fair treatment across different overlap conditions. 
}

\begin{table}[htbp]
\centering
\caption{{Performance of our proposed VUG under different overlap ratios on the Epinions dataset. 
}}
\label{tab:overlap_ratio}
\begin{tabular}{lcc|cc}
\toprule
\multirow{2}{*}{\textbf{Overlap}} & \multicolumn{2}{c|}{\textbf{NDCG@10}} & \multicolumn{2}{c}{\textbf{UGF@10}} \\ 
\cmidrule(lr){2-3} \cmidrule(lr){4-5} 
 & w/o & w/ & w/o & w/ \\ 
\midrule
25\%  & 0.0527 & \textbf{0.0592} & 0.0062 & \textbf{0.0037} \\ 
50\%  & 0.0552 & \textbf{0.0562} & 0.0102 & \textbf{0.0097} \\ 
75\%  & 0.0502 & \textbf{0.0520} & 0.0072 & \textbf{0.0002} \\ 
100\% & 0.0583 & \textbf{0.0700} & 0.0026 & \textbf{0.0001} \\ 
\midrule
\multirow{2}{*}{\textbf{Overlap}} & \multicolumn{2}{c|}{\textbf{NDCG@20}} & \multicolumn{2}{c}{\textbf{UGF@20}} \\ 
\cmidrule(lr){2-3} \cmidrule(lr){4-5} 
 & w/o & w/ & w/o & w/ \\ 
\midrule
25\%  & 0.0634 & \textbf{0.0711} & 0.0074 & \textbf{0.0010} \\ 
50\%  & 0.0658 & \textbf{0.0689} & 0.0118 & \textbf{0.0065} \\ 
75\%  & 0.0612 & \textbf{0.0637} & 0.0050 & \textbf{0.0004} \\ 
100\% & 0.0589 & \textbf{0.0721} & 0.0029 & \textbf{0.0018} \\ 
\bottomrule
\end{tabular}%
\vspace{-1mm}
\end{table}

\subsection{Efficiency Analysis}
\label{sec:efficiency_analysis}
\begin{figure}[htbp]
    \centering
    \includegraphics[width=1\linewidth]{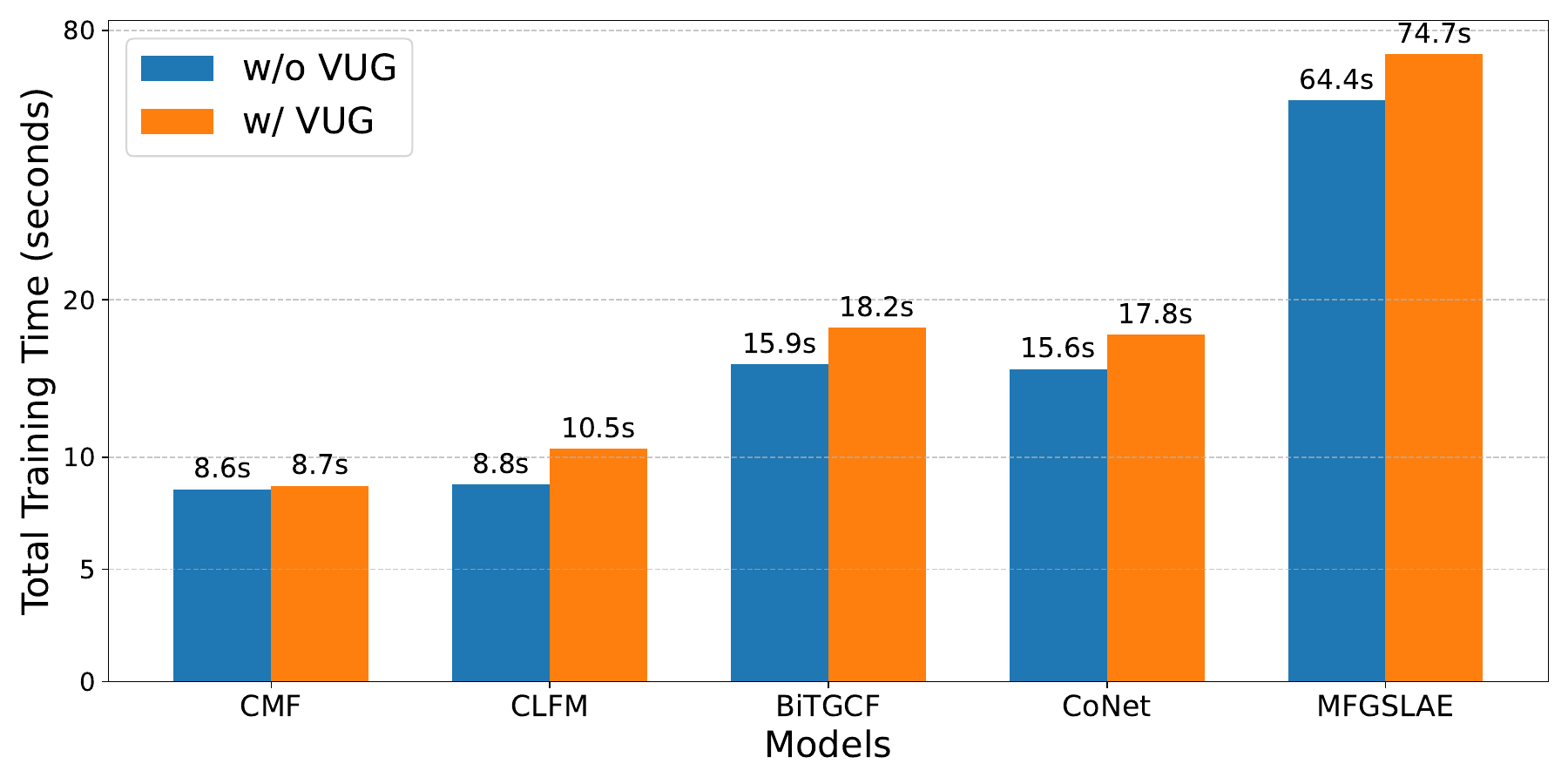}
    \caption{Training time of CDR models on the Epinions dataset, for comparison with and without VUG.}
    \label{fig:efficiency}
\end{figure}

This section analyzes the computational overhead introduced by VUG.  Figure~\ref{fig:efficiency} presents the additional time cost incurred by incorporating VUG into the backbone method.  As shown, VUG introduces negligible overhead. This efficiency stems from our simple virtual user generation strategy, which avoids complex computations.  Because VUG provides performance gains with minimal additional computational cost and effectively mitigates the unfairness arising from overlapping and non-overlapping users, we believe it has the potential to be used in real-world applications.

\section{Related Work}

\subsection{Cross-Domain Recommendation}
Cross-domain recommendation (CDR) aims to leverage information from a source domain to enhance the accuracy of the recommendation in a target domain. In this study, we focus on CDR scenarios where there are overlapping users across domains. 

For instance, CMF~\cite{SG08} was among the first to pioneer the joint factorization of multiple rating matrices by sharing latent parameters across domains. CoNet~\cite{HZY18} introduces cross-connection units to facilitate the transfer of useful knowledge between domains. KerKT~\cite{ZLW18} employs kernel-based methods and domain adaptation techniques to align user features across domains. 
DARec~\cite{YYB19} adopts an adversarial learning perspective to extract preference patterns for overlapping users, enhancing cross-domain knowledge transfer.
BiTGCF~\cite{LLL20} not only exploits high-order connectivity within the user-item graph of a single domain but also facilitates knowledge transfer across domains by leveraging overlapping users as bridges. 
TMCDR~\cite{ZGZ21} uses task-oriented meta network to transform the user embedding in the source domain to the target domain after the pretraining stage.
VDEA~\cite{LZS22} utilizes dual variational autoencoders to achieve both local and global embedding alignment, thereby capturing domain-invariant user representations.
CDRIB~\cite{CSC22} utilizes the information bottleneck principle to decide what information needs to be shared across domains.
~\cite{ZLP23} uses the anchor users in various domains as the learnable parameters to learn the task-relevant cross-domain correlations.
COAST~\cite{ZZH23} introduces a cross-domain heterogeneous graph to capture high-order similarity and interest invariance across domains by unsupervised and semantic signals. 
MACDR~\cite{WYW24} explores the utilization of non-overlapping users in an unsupervised manner, broadening the scope of cross-domain recommendations. 
MFGSLAE~\cite{WYZ25} designs a factor selection module with a bootstrapping mechanism to identify domain-specific preferences and transfer shared information.



\subsection{Fairness in Recommendation}

Extensive studies have highlighted that recommender systems can perpetuate and amplify biases, resulting in unfair treatment across user groups~\cite{chen2025investigating, wu2024bsl, yoo2024ensuring, xu2024fairsync, chen2023improving, zhang2023fairlisa}. 
In response, researchers have proposed various fairness definitions, including individual fairness~\cite{biega2018equity}, envy-free fairness~\cite{EnvyFree}, counterfactual fairness~\cite{DBLP:conf/nips/KusnerLRS17, PCFR}, and group fairness~\cite{zemel2013learning}.
Among these, group fairness has emerged as a central theme due to its intuitive interpretation and direct focus on addressing disparities among different user groups in terms of recommendation distributions or performance metrics~\cite{lifairness, wang2022survey}. 
Key instances of group fairness include demographic parity~\cite{kamishima2011fairness, FairGo}, which ensures similar treatment for different groups. 
Notably, \emph{UGF}~\cite{UGF} has been recognized as a generic group fairness metric in the recommendation literature, effectively representing the equal opportunity principle by measuring disparities among user groups in recommendation quality.
To enforce group fairness, a broad spectrum of methods has been adopted, such as regularization-based approaches~\cite{FOCF, togashi2024scalable, shao2024average}, adversarial learning~\cite{bose2019compositional, PCFR, yang2024distributional}, and re-ranking techniques~\cite{UGF, xu2023p}. 

Recently, more specialized fairness challenges have drawn increasing attention as different recommendation scenarios exhibit diverse unfairness characteristics~\cite{liu2024faircrs, chen2025causality, lin2022towards, DBLP:conf/sigir/WuXZZ0ZL022, liu2022fairness}.
For example, \cite{liu2024faircrs} investigates conversational recommender systems (CRSs), revealing that the inherently notable long-tail phenomenon can lead to disparate treatment among user groups with different interaction levels. 
In multimodal recommender systems, \cite{chen2025causality} identifies the sensitive attribute leakage in different modalities, and proposes to disentangle such sensitive information in user modeling. \cite{tang2024fairness} first highlights fairness concerns within CDR, specifically addressing sensitive attribute bias and proposing a data reweighting strategy.

Different from these works, we focus on another pivotal aspect of CDR unfairness phenomenon: the disparity between overlapping and non-overlapping users. We confront this challenge directly by generating \emph{virtual user representations} in the underexplored domain for the otherwise non-overlapping users, enabling them to more fully profit from cross-domain learning and thus narrowing the fairness gap in CDR.
\vspace{-1mm}





\vspace{3mm}
\section{CONCLUSION}
In this paper, we identify and address a critical unfairness issue in cross-domain recommendation systems: overlapping users disproportionately benefit from knowledge transfer, while non-overlapping users receive minimal advantages or even experience performance degradation. To mitigate this disparity, we propose a novel approach that generates virtual users in the source domain to represent non-overlapping users. Our method comprises two key components: a generator that synthesizes virtual users and a limiter that ensures these virtual representations accurately capture the characteristics of the source domain.

By promoting equitable outcomes, we believe that our approach can help not only enhance user trust and satisfaction but also foster a more inclusive recommendation ecosystem. In real-world applications such as e-commerce and media streaming, reducing bias for non-overlapping users can create a fairer user experience, ultimately benefiting both users and service providers. In the future, we will explore whether unfair phenomenon exists in other CDR methods, such as review-based approaches~\cite{LiuZH022, FuPWXL19}, and investigate the potential benefits of applying our method VUG. We also plan to actively explore the application and deployment of VUG within real-world industrial scenarios.
\begin{acks}
This work is supported by Hong Kong Baptist University IG-FNRA Project (RC-FNRA-IG/21-22/SCI/01), Key Research Partnership Scheme (KRPS/23-24/02), NSFC/RGC Joint Research Scheme (N\_HKBU214/24), and National Natural Science Foundation of China (62461160311).
\end{acks}

\clearpage

\balance
\bibliographystyle{ACM-Reference-Format}
\bibliography{paper}

\end{document}